# Private Power and Public Interests:

An Ethnographic Examination of the Power Outages in Texas in February 2021.

**Authors: William P. Wagner IV, Siyu Xiang, Chien-Ting (Ted) Chang**

***Abstract:*** *In 21$^{st}$ century America, to many observers, the idea that 10's of millions of Americans could lose power and heat for multiple days in the middle of a record cold snap, was unthinkable. It came as an even greater surprise that it would be Texas – arguably one of the world's energy capitals – that failed to provide sufficient power to its residents. This paper explores the events that led to the outage, the experiences of those who lived through it, and the situation in Texas one to two months after the event. We have taken an ethnographic approach to capture both the empirical aspects of the situation, and the more interpretive descriptions of the accounts and thoughts of the participants. We believe this ethnography of events in Texas can serve as foundational evidence and therefore can be generalized to a wide variety of situations and methodologies.*

## INTRODUCTION

When forecasters warned of a cold snap that would engulf most of the United States, the focus of much media attention immediately turned to the Northeast, where feet of snow were expected to accumulate and shut down that region of the country [1][2]. The storm was expected to roll across the plains from the west. In California, residents were warned of severe winter rains, and in the Northwest, harsh winter conditions were predicted. Colorado's ski resorts looked forward to a great base of snow for its mountain resorts. The South faced extreme cold. The country prepared for what was thought would amount to just more indoor activity after almost a year of sequestering because of Covid-19. There was little concern that the cold temperatures in Texas would cause power outages that would prove fatal for more than 80 people – more than half of whom would die of hypothermia.

The philosophy of our investigation is grounded in what Peter Godfrey-Smith describes as Scientific Realism [3]:

> ***Common-sense Realism Naturalized:***
> *We all inhabit a common reality, which has a structure that exists independently of what people think and say about it, except insofar as reality is comprised of thoughts, theories, and other symbols, and except insofar as reality is dependent on thoughts, theories, and other symbols in ways that might be uncovered by science.*
>
> ***Scientific Realism:***
> *1. Common-sense realism naturalized.*
> *2. One actual and reasonable aim of science is to give us accurate descriptions (and other representations) of what reality is like. This project includes giving us accurate representations of aspects of reality that are unobservable.*
> *-p.176*

Within the context of a mixed-methodology qualitative approach, we have applied a Critical Realist and somewhat hermeneutic methodology [4][5], as well as taking an ethnographic [6][7] approach to the reporting. Zachariadis et al. cite Mingers' ontology of Critical Realism (CR) as "… opening up a particular methodological space that lies between empiricism and interpretivism (Mingers)." [8]. We believe we have successfully combined the methodological considerations of CR with an engaging ethnographic presentation of the results.

**Research Questions**

**RQ1**: How does media coverage of a natural disaster and conveyance of relevant information qualitatively compare with the conversations in the Twittersphere?

**RQ2**: Can we make any philosophically sound observations about the conveyance of emergency information and/or the structure of emergency information systems?

**RQ3**: Are there methodological implications for qualitative research?

Specifically, we examine the events of mid-February 2021, in Texas. This short timeframe provides a window we believe can be generalized, yet it is specific enough that we can provide the requisite historical and philosophical context to provide ethnographically thick context and value.

We use coding to help identify themes within the dissemination of information with regards to the events in Texas. This type of examination mirrors Kotlarky et al. [9] in their examination of knowledge sharing in the offshore-outsourcing process, who describe their codification approach as "an appropriate arena in which to ground these concerns."

The events in Texas provide a unique comparative opportunity because the Texas electrical grid is mostly disconnected from the rest of the U.S. power grid and is governed by an entity called the Electric Reliability Council of Texas (ERCOT).

Within this context we examine how the various players are portrayed and how they are seen both in media and through the eyes of Twitter users. We hope to capture sufficiently thick description that any reader can draw conclusions about the situation that will hopefully lead to an understanding of the interpretations that Twitter users draw, as well as empirical evidence of an interpretivist reality – Critical Realism.

# HISTORICITY

## *Spanish Colonialism*

During the 19th century, Texas was the site of much conflict. Europeans and American immigrants carved the area known as Texas from lands previously inhabited by Native American tribes like the Apaches and Comanches in the northern portions, and the Mexican Government in the south.

While any justification for the European/American migration to Texas is beyond the scope of this paper, we can empirically say that there is a long tradition of anyone living in the area now known as Texas having to fight violently to protect their own interests – be they Native Americans, Mexicans, European Catholic immigrants, or American immigrants.

Mexico gained its independence from Spanish colonialism in 1821. The Spanish had used the two northern Mexican states of Coahuila and Texas as a buffer between the North American Native tribes like the Apache, Comanche, and Navajo, and the southern areas of Mexico. At the time, there were far more Native American tribes than anyone else on the land, and they had proved both capable and willing to use force to defend their lands against Mexican colonialism.

The Mexicans living on the land were known as Vaqueros. They were the original cowboys. They were cattle ranchers, and their lifestyle of large cattle ranches perhaps conflicted with the Native American traditions of a more nomadic culture.

In late 1820, the Spanish government signed an agreement with Moses Austin, who subsequently died. But, Stephen Austin, upon his father's death completed the agreement in 1821 for amongst other things, the arrival of 300 Catholic families and their possessions (including slaves) and land. However, this agreement was declared void when Mexico declared independence from Spain later that year.

Instead, the newly formed Mexican government chose to pass a series of State and Federal laws. Amongst the provisions of the National Colonization Law [10][11] was a provision that allowed settlers in the Mexican state of Texas to have slaves, even though slavery had been declared illegal when Mexico revolted, and the rest of Mexico would formally outlaw slavery in 1829. And they made Stephen Austin "Empresario of Texas".

E. R. Schlereth, et al. take an ethnographic approach in their post-modern analysis of the Texas Revolution [12] that attempts to present the events from both the perspective of the Mexican government and the perspective of the Texicans, as well as a final essay describing some of the effects of the Revolution, and a valuable re-examination of the debate around Texas's admittance to the United States. We will utilize a similar approach to briefly frame what we believe to be important political context when interpreting the Tweets and media coverage of the power outages in 2021 Texas.

## *Texas Revolution*

While many Americans think of Texas as having always been a US state, it was originally a Mexican state. And the newly formed in 1821 Mexican government strongly encouraged Europeans and Americans to settle in Texas. Many authors agree that the intent of the Mexican government was to incentivize the Catholic, European, and American immigrants to occupy lands in Texas so that they would fight the Native American tribes on behalf of the Mexicans.

As Europeans began to arrive in greater numbers, the Mexican government then imposed tariffs, restrictions on European settlement in Texas, and other measures designed to enforce Mexican rule over their state of Texas. Per the Texas State Historical Association [13], the Texas Revolution was fought between 1835-1836, out of which emerged the Republic of Texas; however, Britannica [14] documents a series of massacres from 1826 through the Texas Revolution 10 years later, by all sides in an attempt to assert violent control over a given area. From house-to-house intimidation by the Texans, to armed skirmishes where the Texans and the Mexicans fought together to prevent other immigrants from arriving, and finally culminating in the Battles of Coleto and Béxar – two key points required for travel between northern and southern Texas.

400 European/American prisoners were massacred in the prison at Golita after being captured at Coleto. And Béxar was later re-named San Antonio, where 180-189 Texans were killed when they chose to make their fateful stand in a mission known as The Alamo.

After these two battles, the Texans pulled back towards the north, until given an opportunity to attack General Santa Ana's troops while they were resting on a road. 800-900 Texans were able to capture or kill approximately 1100 Mexican troops, while reportedly only taking 8 casualties. And General Santa Ana, himself was captured.

The Texas Declaration of Independence was framed and issued by the Convention of 1836 at Washington-on-the-Brazos [15]. But while he had been fighting the campaign in Texas, Santa Ana had been deposed as leader of his own government.

Eventually an agreement (actually two agreements – a public and a private one) was signed between Santa Anna and the Texans to end the Revolution, but that agreement continued to be contested by the Mexican government for decades to come.

## *"Manifest Destiny"*

At the same time, there was a period of hundreds of years from when Europeans first began expanding westward that culminated in the American "Manifest Destiny." Harvard Professor Frederick Merk described the conflict in delicate terms in 1966 [16] when he said, "Expansionism is an exciting study. It does not, however, always leave the spirit of the reader uplifted. It involves elbowing owners of property rudely to one side and making away with their possessions." (p.8)

In reality, this "elbowing" manifested itself in Texas in the form of at least two wars and countless dead in other armed conflicts between American Immigrants and Native Americans.

Further, while it is important to recognize that this land was taken by its current occupants, from the previous occupants by force, it is also important to recognize that at the time, this was "the way of the world." International borders were undergoing change all over the world. Former colonies, like the United States, were seeking to establish themselves – and this often involved military confrontations.

There were, of course, long histories and traditions of laws amongst the native residents that the Americans completely disregarded; but these too, often applied only to within a given tribe. There are numerous examples of inter-tribal warfare for the purpose of establishing boundaries for thousands of years prior to the invasions of the Europeans.

But these conflicts had been on a different scale and followed a different set of military restrictions and capabilities. The consensus is the European and American colonists had better technology than the Native tribes, and they were far more willing to exploit the land, technology, and people for expansion than the Native tribes had been.

Since its inception, Texas and Texans have fostered an image of rugged independence, and distrust of federal government – originally a distrust of the Mexican Federal government, then later, the U.S. Federal government.

In 1844, James K. Polk was elected the 11th President of the United States and proclaimed America's "Manifest Destiny." As part of this expansion, the borders of Texas and California were to be formalized between the U.S. and Mexico. The Mexican government refused to accept the U.S.'s proposal.

But Polk needed Texas as a U.S. State in order to fulfill his Manifest Destiny. So, after over a decade of petitioning the United States for admittance, the U.S. Congress finally voted to accept Texas's annexation, and on December 29, 1845, Texas entered the United States as a "Slave State" – a state where slavery was still legal. This would then lead to two more wars.

The first was the Mexican American War. This was the first time the U.S. had fought a war on foreign soil, and it ended with the sacking and occupation of Mexico City and the downfall of the still fragile Mexican government.

In January 1846, Zachary Taylor's army advanced across the Rio Grande River at Corpus Christie. Over the next two years they would fight their way south, while US Naval forces blockaded the Mexican coast. The war ended with the siege and destruction of Mexico City, accompanied by the obvious threat of attack on other cities.

In 1848 the Mexican government signed the Treaty of Guadalupe Hidalgo, in which the United States gained California, Arizona, New Mexico, and the Rio Grande boundary for Texas, as well as portions of Utah, Nevada, and Colorado. This "Treaty" was signed without the consent of any of the Native American tribes that lived in any of those states.

Then, only 12 years later, Texas would vote to secede from the Union, joining the Confederacy, and triggering the U.S. Civil War. Table 1 shows the Texas Secretary of State [17] official statistics for Presidential elections through the first 28 years of Texas's history.

After the U.S. Civil War, Texas was placed under an occupational U.S. government from 1867-1870 during the first years of what would turn out to be a troubled reconstruction. And finally, in 1870, Texas was readmitted to the United States, and participated in the national elections for President two years later.

Whether one believes the Manifest Destiny of expansionism and American "exceptionalism" justified the death and racism that it fostered is a much larger discussion; however, we believe anyone can understand why it is that many Texans felt like they fought and died for the right to live on that land. It was quite literally a warzone for almost 55 years straight - through seven different Federal governments. There were no gun laws, and any other laws were only subject to enforcement if whatever local contingent was in power decided to enforce them. Quite often land ownership came down to the last man standing.

And now, to this day, it is still a topic of conversation in Texas as to whether the state should secede from the Union. In other parts of the United

*Table 1 – First 28 Years of Texas Presidential Election Results*

| Name | Party | Votes |
|---|---|---|
| **1848** | | |
| *Lewis Cass* | Dem | 10,668 |
| *Zachary Taylor* | Whig | 4,509 |
| **1852** | | |
| *Franklin Pierce* | Dem | 13,552 |
| *Winfield Scott* | Whig | 4,995 |
| **1856** | | |
| *James Buchanan* | Dem | 31,169 |
| *Millard Fillmore* | Whig | 15,639 |
| **1860** | | |
| *John Breckinridge* | Dem | 47,548 |
| *John Bell* | Con'l Union | 15,438 |
| **1861 - Texas secedes from Union and joins the Confederate States of America.** **1861-1865 - Texas in the Civil War.** **1867-1870 - Congressional (or Military) Reconstruction.** **1870 - Texas readmitted to Union.** | | |
| **1872** | | |
| *Horace Greeley* | Dem | 66,546 |
| *Ulysses Grant* | Rep | 47,468 |
| *Charles O'Conor* | S-O Dem | 2,580 |
| **1876** | | |
| *Samuel Tilden* | Dem | 104,755 |
| *Rutherford Hayes* | Rep | 44,800 |

States this conversation would be considered ridiculous, and in some places even traitorous, yet in Texas there is currently talk of "Texit." H.B. No. 1359 "A BILL TO BE ENTITLED" would put a referendum on the ballot in 2021 to ask Texans if they would favor secession: "AN ACT relating to proposing a referendum to the people of the State of Texas on the question of whether this state should leave the United States of America and establish an independent republic."

## *Oil Industry*

Texas also has a long history and association with the oil and gas industry, but unbeknownst to the authors before starting the investigation was the extent to which oil was used for medicinal uses, as well as for lubrication, long before refining allowed its use as a source of power. Rister's [18] seminal analysis of the early days of the oil industry, published in 1949, and interestingly paid for by the Standard Oil Company via a grant to the University of Texas, includes the following extensive account (p.33), and it bears inclusion in its whole:

> *The use of petroleum is frequently mentioned in the writings of the ancients, but the earliest documentary account of its use in North America dates from a blustery July day in 1543, when storm-tossed Spanish ships - bearing the survivors of the ill-fated De Soto expedition on their way back to Mexico — sought shelter on the Gulf coast of Texas near Sabine Pass.*
>
> *The record of this expedition, published at Evora, Portugal, fourteen years later, tells how "The vessels came together in a creek, where lay the two brigantines that preceded them." On the water, floating about the ships, was a dark scum, which the Spaniards called "cope." Since it was like the pitch which they had used in Spain to calk their ships, they "payed the bottoms of their vessels with it."1*
>
> *This was the white man's first use of a petroleum product in what is now the United States— sixty-four years before the first English colonists stepped ashore at Jamestown. Texas Gulf Coast oilmen of today [1949] confirm this Spanish narrative, for oil springs are yet found near Sabine Pass, and "cope" is occasionally seen upon the water.*
>
> *Indians had used petroleum before the white explorers came. Their traditions include stories of their visits to oil springs in times of affliction; and both early Spanish and Anglo narratives mention similar incidents. The Indians bathed in oil springs to drive away rheumatic pains, applied the oil as an ointment to cuts, burns, and sores, and drank it as medicine.*
>
> *1 Edward Gaylord Bourne (ed.), Narratives of the Career of Hernando de Soto, 1290.*

While California would have the Gold Rush in 1849, Texas and the Southwest would start to see their fortunes explode with "Black Gold" around the same time.  However, the early days of the oil market were markedly different than those after the turn of the century.  In the 1800's oil was primarily valued for its medicinal properties, with a secondary market of lubrication.  Oil was sold by the barrel for lubrication, but the same oil could be priced much higher and sold by the ounce for medication.

Then, towards the end of the century, improvements in basic refining and basic drilling techniques saw a dramatic shift in the oil industry.  Oil became a source of power.  And suddenly the challenge became not one of retrieving the oil from the ground, but instead - storage.  Wagons and trains were insufficient means of distribution.  As the oil fields of Beaumont, LA came online, the price of a barrel of oil dropped to $.03 whereas a glass of clean water was $.05.

Although extremely dirty and dangerous, this was cheap and bountiful power, and the desire to sink a well while a stake was producing led to many entities literally drilling through existing structures (Figure. 1) [19], with the consent of their owners.  Rister cites examples of churches, banks, stores, and individual properties that gladly allowed a derrick through their structure for a share of the returns.

Tens of thousands of individuals and families raced to the southwest of the U.S. in search of fortunes in either oil, cattle, or any of the myriad service, support, and sometimes predatory industries that sprung up.

As the world moved into the industrial age and became dependent on fossil fuels to power progress, oil drilling platforms became commonplace in Texas, Louisiana, and Oklahoma at the end of the 19th Century.  Some companies would even leave a well uncapped for the tourists and publicity photos of the "Gushers."  This quickly became an issue, both from a safety standpoint, and from a financial concern.  Large companies came into existence and went bankrupt as huge sums of money were spent, only to have a field dry up.

It was also around the turn of the century that oil well workers and owners alike came together to demand some regulation from the government.  Well fires were becoming commonplace and because the wells were so close together, they often spread from one to another.  Laws were introduced that limited the number of wells within a given area, and also prevented the sale of alcohol within 100 feet of a well [11].

Drilling for oil was a dirty and invasive process, and it was not efficient.  That would change over the next 50 years.  Rister points out in his 1949 publication that more oil was extracted in the first half of 1949 than all previous production combined.  The early oil business was dominated by "Wildcatters" – individuals willing to dig a hole 300 feet deep by hand or risk their own fortunes on a derrick.  And even as large corporations moved in, the oil and gas business has remained a mix of large entities dominating the landscape, with many wildcatters and small operators still hoping to make their fortunes from Black Gold.

## ERCOT

And within this context of independence, modern Texas has also chosen a different approach to providing power to its residents.  Whereas most of the rest of the country is connected by overlapping power grids that borrow and loan electricity to each other to offset surplus or deficits

*Figure 1 - Oil Derrick Drilled Through a Store*

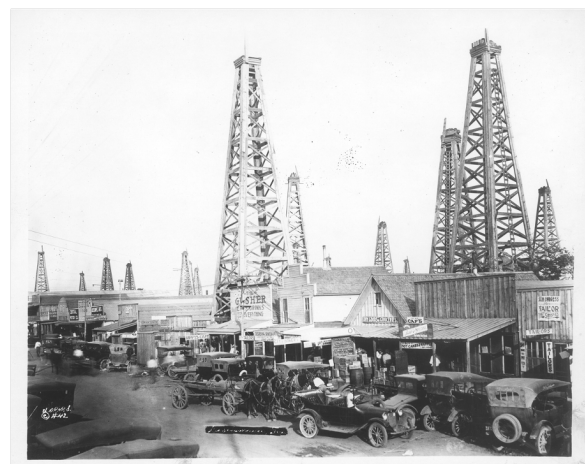

in power, most of Texas is independent from the US power grid.

Most Texans receive their power from a range of local providers in Texas that theoretically compete to provide the best experience for the customer. Instead of a government directly mandating prices or standards of service, consumers in Texas are offered power from multiple private electricity companies that are part of The Electric Reliability Council of Texas (ERCOT) [20].

> *The Electric Reliability Council of Texas (ERCOT) manages the flow of electric power to more than 26 million Texas customers -- representing about 90 percent of the state's electric load. As the independent system operator for the region, ERCOT schedules power on an electric grid that connects more than 46,500 miles of transmission lines and 680+ generation units. It also performs financial settlement for the competitive wholesale bulk-power market and administers retail switching for 8 million premises in competitive choice areas. ERCOT is a membership-based 501(c)(4) nonprofit corporation, governed by a board of directors and subject to oversight by the Public Utility Commission of Texas and the Texas Legislature. Its members include consumers, cooperatives, generators, power marketers, retail electric providers, investor-owned electric utilities, transmission and distribution providers and municipally owned electric utilities.*
>
> *- Mar 28, 2021*
> *https://www.ercot.com*

Adding another dimension to the situation, ERCOT allows its consumers to purchase electricity via two different pricing mechanisms. One is an electric service agreement whereby consumers buy their electricity at a normalized price. This price is usually determined as an average of the previous year's prices. Under this plan, consumers may pay slightly more per KwH than the wholesale rate during times of low usage but see greater savings during times of high demand and there are no "spikes" in prices.

A second plan offers consumers "wholesale" electricity prices, based on the laws of supply and demand. When there is surplus, the consumers are supposed to see lower prices, and when the demand grows, the consumers are supposed to pay a higher rate.

And finally, there is the transition from oil and gas as the primary suppliers of the world's energy to a focus on sustainable resources. Texas has long been amongst the world's leading producers of oil and gas and there is significant power and money to be had in the control of that production. As such, there is natural reluctance on the part of many established interests in Texas to shift from oil and gas to wind and solar. However, this shift has begun, as acknowledged in the Executive Summary in ERCOT's *Report on the Capacity, Demand and Reserves (CDR) in the ERCOT Region, 2021-2030, December 16, 2020*[21]:

> *New generation resources, including a significant amount of utility-scale solar, continue to be added to the ERCOT region at a rapid pace, resulting in higher planning reserve margins over the next several years…*
>
> *ERCOT has seen a significant increase in utility-scale solar resources and based on the grid operator's current inter-connection queue for new generation projects, this trend is expected to continue over the next several years.*
>
> *The grid operator is also seeing continued and accelerated growth in rooftop solar projects. In response, ERCOT included its first, separate rooftop solar PV forecast in the CDR. The forecast was created to show the incremental capacity growth beyond the historical growth trend reflected in the load forecast.*

Texas lies on the $31^{st}$ parallel North, so it sees temperatures that may fluctuate between freezing during the winters, followed by long months of warmth and heat. The Texas panhandle is well known for its sprawling prairies, rolling tumbleweeds, and armadillos sunning on the road.

As such, the vast majority of their wind and solar power is tailored to the warmer months. The distribution network attempts to take this into account, per ERCOT, during non-winter months, wind and solar should account for upwards of 30% of the power generated in the state. During winter months, wind and solar are supposed to provide approximately 8% of the state's power production.

It is interesting to note that in the December 2020 CDR Report (two months before the outage in February) the Executive Summary only addresses the upcoming summer capacities and has no mention of the winter capacity that would soon be overwhelmed.

## Socio-Political Context

While it is beyond the scope of this paper to provide an in-depth ethnographic analysis of the socio-political climate in Texas, we believe there is sufficient empirical support to draw some broad generalizations.

In the U.S., Republican voters are often described as "Red" voters and states, and we think of Democratic voters as "Blue" voters and states. These are arbitrary colors that appear to draw their roots from the maps used in television election coverage starting around 2000. These red/blue generalizations have no political significance, but they provide valuable context when reading Tweets, or analyzing the content contained therein.

Texas is currently considered to be a "Red" state [22] with some distinct pockets in and around the metropolitan areas and along the southern border with Mexico that are heavily "Blue"; and, as described below, some would say that the shifting demographics lean towards Texas becoming more "Purple". Austin was described to one researcher as "A blueberry in the tomato soup." While this perspective provides some surface description, maintaining internal validity with regard to philosophical considerations in ethnography requires us to interpret these descriptions in context.

Texas voted for Donald Trump in 2016 and 2020 [22]. Official Secretary of State totals show that Texas has been reliably "Red" with regards to its presidential votes since 1980, and in fact, Texas has voted Republican in every Presidential election since 1980. Interestingly, prior to 1980, however, Texas had never voted Republican.

We have so far not seen any evidence of either the demographics of the state's "Blue" voters, or the afore-mentioned change from "Red" back to "Purple." For that evidence, we turn to a more detailed examination of voting results at a county level (Figure 2).

Now we see that there are pockets of the state that are distinctly red or blue. But what evidence do we have of a shift towards "Purple?" This is a far more complex question than the previous ones because it involves at least some prediction about the future. Again, a detailed analysis of the political trends of Texas is beyond the scope of this paper, but we can look to demographic shifts that would indicate that if the voting blocs [23] continue to vote the way they have in the last 20 years, then changes in the population will increase the percentage of "Blue" voters and shift the entire state towards "Purple".

This type of analysis is not philosophically or logically sound, because it assumes that one bloc will not change their trend. And as we have seen, Texas used to vote primarily Democrat and shifted dramatically to Republican in 1980. If the same thing happened again by any significant bloc of voters in Texas, we would see political change not based on current demographic data.

*Figure 2: 2020 Presidential Election Results by County*

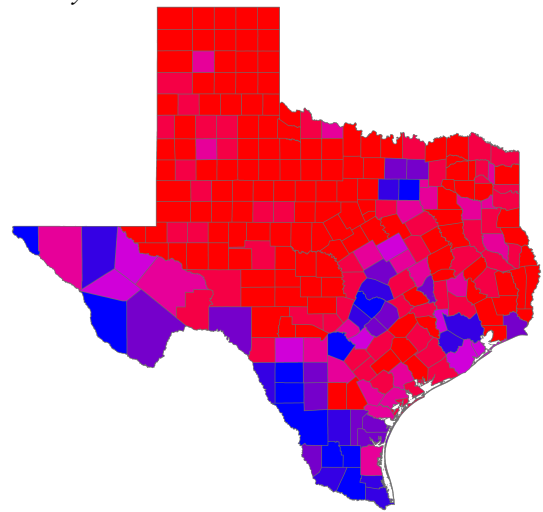

However, it should be noted that based on the historicity presented earlier in this paper, Texan voters as a bloc have shown a propensity over the last 160 years to vote based on nationalistic tendencies with regard to Texas being its own entity, and they tend to exhibit significant inertia with regards to changing parties.

## MEDIA LANDSCAPE

The media landscape in the U.S. is of course extremely varied, but in the best traditions of ethnography we can generalize to abstract some meaningful value.

Traditionally, mainstream media was theoretically held to a high standard of journalistic integrity which created a cycle of trust between the news outlets and their viewership or readership. One interviewee phrased it as follows:

*I don't know what is mainstream media, but I think of it as a more trustworthy organization.*

However, the media landscape has changed considerably in the last decade with a shift in mid-2016-2017 when more people started reporting for the first time that they preferred to get their news from social media than print.

We will show that this idea of "trust" cannot be considered a valid method of distinguishing the groups without introducing ideological considerations. However, we use a lens of *trust in the source* as a means to examine the societal (interpretivist) relationships that evince themselves in the Tweets, interviews, and media coverage that we have studied.

Our primary distinction was to start by separating mainstream media from those outlets with a smaller viewership. This distinction allows us to distinguish the various sources based on an estimated viewership count, rather than along any kind of ideological division. Finding official data about viewership is difficult from a peer-reviewed primary source standpoint, but according to publicly available information from statista.com [24] and Neilsen [25], the biggest three cable news outlets are FOXNews™, CNN™, and MSNBC™.

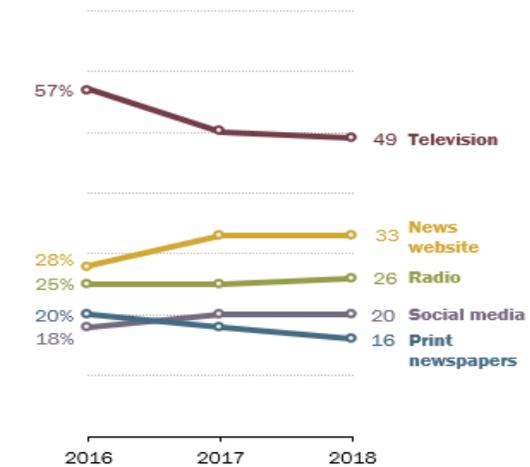

*Figure 3 - The shift from Print to Social Media*

For purposes of this paper, mainstream media can be considered to include news outlets like AP™, Reuters™, UPI™, FOXNews™, The New York Times™, The Wall Street Journal™, CNN™, ABC™, MSNBC™, The Guardian™, BBC™, and in a significant change since the 2020 elections, Newsmax™ might be considered amongst that group. Other entities that might be considered mainstream by some would be outlets like Breitbart™, Al Jazeera™, One America News Network™ (OANN), NPR™, Vox™, ARS Technica™, and others who have a group of followers who may also overlap with the other groups above. Many readers will be quick to point out that there are many more outlets in that list that might be described as "left leaning" or "liberal media" than there are "conservative" outlets. That reflects the media landscape in general. While FOX news is by far the largest single provider, and accounts for a significant portion of news both overall, and on the "right", the media landscape on the "left" is characterized by a more divided network offering.

This has changed since the 2020 Presidential election, and "The Big Lie[1]." In what is a revelatory change, FOXNews lost viewership to rivals Newsmax and OANN, because FOX had called Arizona in favor of Biden on the evening of the election.

The viewers were not interested in a factual accounting of the election, they were looking for ideological reinforcement for their support in Trump. Fox would then respond by adding more "Opinion" shows to their primetime lineup, removing "News" programs, and attempted to cater to the premise of a stolen election. So much so, that as of this writing, they are being sued by two election balloting companies for over $2.5 billion for defamation.

The number of outlets of a given political leaning should not be an indication of bias with regards to our research, as we have tried to collect sufficient empirical evidence regarding the events, in conjunction with an in-depth analysis of the media reporting. We have tried to place the events in their ethnographic context so that we can then evaluate the reporting from the media and from the Tweets. Our goal is to build a sufficiently complete picture of the events in Texas that it withstands generalization.

A second philosophical distinction is between the mainstream media outlets above, and what is commonly referred to as "online" or "social media platforms". This distinction is blurred because the mainstream media outlets listed above usually have significant online presences both independently of social media in the form of web sites (www.some_entity.com) as well as significant presences on many of the social media platforms (@some_entity, #Some_Entity).

Social Media can be thought to include platforms like Twitter™, Facebook™, Snapchat™, Instagram™, TikTok™, and many others. These companies do not create news content themselves, rather they offer a "platform" for their users to create and surface content.

Further, we can make an ethnographic distinction between what might be described as the socio-political identities of the various outlets and their viewership demographic. For example, Fox news viewers are far more likely to be "Red" voters and supporters of Donald Trump than are viewers of CNN or MSNBC.

We can link those three philosophical pieces together with an investigation of possible origins of individual Tweets and the ideas expressed therein (threads) that is grounded in the empirical evidence of the historical and socio-political contexts described earlier in this document.

There has been considerable investigation of the media landscape in the U.S., and we don't want to wade into a methodological discussion of those analyses – that would be a criticalist approach. The graphic (Figure 1) [26] from Vanessa Otero's whitepaper [27] is one illustration of the media landscape and tries to account for some political bias as well as trying to account for the veracity of the given outlet. While her methodology is not peer reviewed, we found it sufficiently rigorous to be beneficial in at least an ethnographic context. In other words, it is secondary or even tertiary evidence of the actual landscape, but it is primary evidence with regard to understanding how various people are looking at the landscape.

# SOCIAL MEDIA INVESTIGATION

## Internal Validity and Methodology

Qualitative methodology allows for a wide variety of techniques for analysis; however, the underlying philosophical foundations must remain consistent. Myers [28] p.30 states that," every qualitative researcher should make their philosophical assumptions explicit." Myers puts those assumptions in three categories: positivist, interpretivist, and criticalist.

---

[1] "The Big Lie" refers to the effort by some to claim that Donald Trump won the 2020 election and had it stolen via nefarious means. This has been repeatedly proven to be false, and those espousing this theory are said to be continuing "The Big Lie."

Whereas a positivist approach would attempt to suggest a "best" solution or an "objective" reality, and a criticalist approach would demand an examination of our methodology in the context of other approaches, an interpretivist philosophical perspective lends itself to the investigation of social media based on the socio-political relationships therein.

Following in the anthropological traditions of Geertz [4], Kuhn [29], Crapanzano [30] and others, this paper assumes political and power relationships to be fundamental to an understanding of the situation. Power can be interpreted in a variety of ways, and we are not attempting to articulate specific consequences of political decision-making or political dealing, other than to help explain or understand any mention of the ideas in the Tweets or in the news reports.

Within this framework of the interpretivist philosophy, we have chosen an inductive grounded theory methodology [31][32], implementing convenience sampling for the interviews, snowball sampling for data collection, and purposeful sampling for final comparison.

We are interested in understanding how the various entities expressed themselves in the context of social media, so we used coding (both Open and Axial) to extract key themes from the Tweets. And finally, Theoretical coding allowed us to extract the ideas of Trust and Time as commodities.

We used the geo information from Twitter to place the Tweets in physical space where possible. And since each Tweet has a timestamp, we were able to track the information in 4-D space.

*Figure 4 – Media Bias Chart*

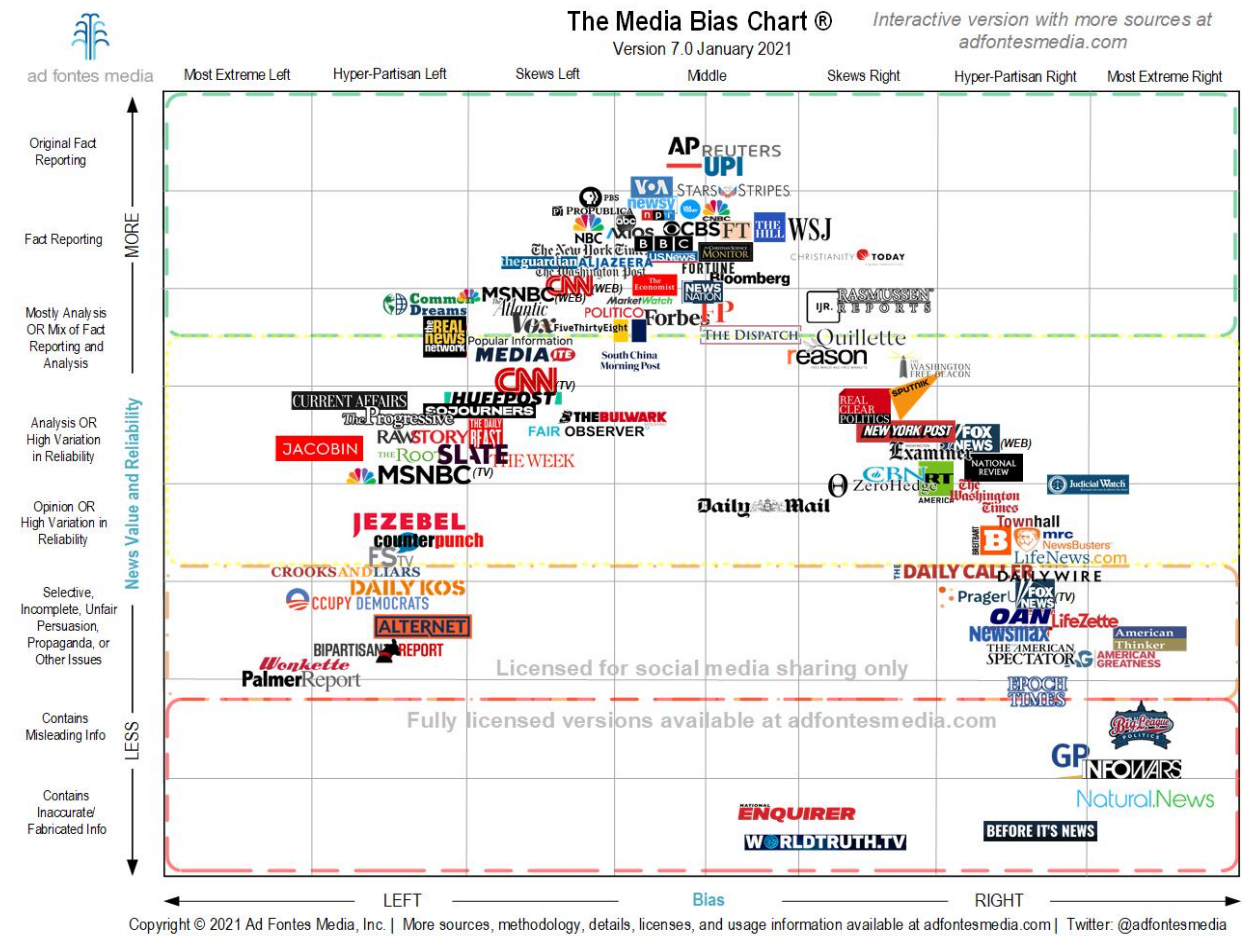

Finally, we place the information in an ethnographic context to allow for generalization from the specific – in this case the week of the power outages in Texas. As such, it was necessary for us to gain some domain expertise of the events, much of which has been presented above, and then interpret the Tweets within our understanding.

We also explore if there is a spatial element of the Twitter conversations that corresponds to the spatial disparities, we see in the Presidential election voting. This disparity presents itself in the media landscape and in the context of the interpretivist philosophy, so we believe it may be fundamental to understanding any given Tweet's origin and sentiment.

Per Myers [28], in his discussion on p.137 about trying to analyze social media, one issue facing the qualitative researcher is the volume of information and data that needs to be processed. To address this concern, we have chosen to focus our attention on the events in Texas, however we used a semi-structured methodology to our interviews that allowed us the flexibility to ask interviewees about their understanding of news reporting in general if they were not familiar with the specifics of Texas. We believe this to be beneficial in that while most of the data gathered was specific to Texas, the additional interviews provide good context to allow us to generalize the findings.

*Twitter™*

We have used Twitter's API via their Academic Research Portal [33] to download 482 Tweets. The Tweets primarily focus on the events of the week of the power outages in Texas, with some padding before, and additional Tweets gathered (convenience sampling) after the events to help the investigation. The initial API call/query used to get Tweets for initial analysis and data exploration is shown in Figure 2. From this query, we received an initial corpus of 482 Tweets. We then used Python to parse the JSON information. Because the query had a requirement of (`has:geo`)

*Figure 5*

```
url = "https://api.twitter.com/2/tweets/search/all?query=(\"electricity\" OR \"power outage\" OR \"ERCOT\") has:mentions (has:geo)&start_time=2021-02-13T00:00:00.00Z&end_time=2021-02-22T00:00:00.00Z&max_results=500&tweet.fields=public_metrics,referenced_tweets,reply_settings,source,text,geo,withheld&place.fields=contained_within,country,country_code,full_name,geo,id,name,place_type"
```

- all these Tweets had a Twitter `place_id` associated with them. This id refers to a specific place in Twitter's database of locations.

Location data is an "opt in" option for Twitter users. By default, no location data is stored or collected regarding individuals or their Tweets. However, users can opt in to share both their location, and/or the location of individual Tweets. According to the Twitter API Documentation [33]:

*Introduction*
*Tweets can be associated with a location, generating a Tweet that has been 'geo-tagged.' Tweet locations can be assigned by using the Twitter user-interface or when posting a Tweet using the API. Tweet locations can be an exact 'point' location, or a Twitter Place with a 'bounding box' that describes a larger area ranging from a venue to an entire region.*

*There are two 'root-level' JSON objects used to describe the location associated with a Tweet: coordinates and place.*

*{*
  *"coordinates": {},*
  *"place": {}*
*}*

*The place object is always present when a Tweet is geo-tagged, while the co-ordinates object is only present (non-null) when the Tweet is assigned an exact location. If an exact location is provided,*

*the coordinates object will provide a [long, lat] array with the geographical coordinates, and a Twitter Place that corresponds to that location will be assigned.*

*Place object*
*Places are specific, named locations with corresponding geo coordinates. When users decide to assign a location to their Tweet, they are presented with a list of candidate Twitter Places. When using the API to post a Tweet, a Twitter Place can be attached by specifying a place_id when posting the Tweet. Tweets associated with Places are not necessarily issued from that location but could also potentially be about that location.*

The Twitter definition of "Place" makes an important philosophical distinction with regard to our analysis: *Tweets associated with Places are not necessarily issued from that location but could also potentially be about that location.*

This essentially means that without further analysis of a given Tweet, it must be presumed to be at best secondary information with regards to a specific event or statement, and quite often it will be tertiary with regards to an idea. So again, while of little to no logical or philosophical importance with regard to understanding a specific historical event, it *can* serve as a primary source for an empirical examination of the ways that Twitter users talk about something.

While we consider this to be a qualitative analysis, we have employed some analysis that might be thought of as quantitative in nature. In his 2005 Annual Review of Science and Technology article [34], Chowdhury defines "Natural Language Processing (NLP) [as] an area of research and application that explores how computers can be used to understand and manipulate natural language text or speech to do useful things."

In our initial investigation of the data two types of NLP were employed. We used the Natural Language Tool Kit (NLTK) library [35] in conjunction with the Valence Aware Dictionary for sEntiment Reasoning (VADER) [36][37] sentiment analysis library; and, for a second method and metric, we used Google's Cloud Natural Language API [38].

None of these tools required us to actually train and build a model – which was not the focus of this particular pilot study. We were able to use a readily available corpus of terms, emojis, stop lists, and go lists, or in the case of Google, we simply accessed their API and took advantage of their most recent iteration of their pre-trained AutoML.

**NLTK + VADER**

The two methods allowed us to quickly examine the 482 Tweets for some basic trends in sentiment. In our examination, VADER was able to understand the sentiment of the Tweets very well, but it was unable to understand to which entity the sentiment should be attributed. In other words, in examining a given Tweet, VADER would be able to say whether someone felt strongly about something, but it could not say what.

The amount of that sentiment can be reported by VADER as four real numbers in the range of -1.0 to 1.0: Negative, Positive, Neutral, and a Compound score (Table 2).

*Table 2 – VADER Sentiment Analysis Sample*

| Tweet | compound | neg | neutral | pos |
|---|---|---|---|---|
| 1 | -0.5719 | 0.072 | 0.928 | 0 |
| 2 | 0.0258 | 0 | 0.977 | 0.023 |
| 3 | 0.4466 | 0.108 | 0.735 | 0.158 |
| 4 | 0.0772 | 0.089 | 0.815 | 0.095 |
| 5 | -0.4404 | 0.172 | 0.828 | 0 |
| 6 | 0.1027 | 0.107 | 0.772 | 0.121 |
| 7 | 0.1531 | 0.1 | 0.768 | 0.132 |
| 8 | 0 | 0 | 1 | 0 |

**Google™**

The Google API allows for multiple methods of NLP within the context of their framework. Google allows for Entity analysis, Sentiment analysis, and a combination of the two called Entity Sentiment analysis, where the NLP attempts to identify sentiment associated with a specific entity.

Again, in our tests, the NLP performed well and was indeed able to report interesting information on a per-Tweet basis. Google's

*Table 4 – Google Sentiment Analysis*

| |
|---|
| Document sentiment score: -0.4000000059604645 |
| Document sentiment magnitude: 1.2999999523162842 |
| Sentence text: @RollinOliver @AJ2theWorld @snbfg4 @davidmweissman @tedcruz They offer those things but at the expense of other things in the environment. |
| Sentence sentiment score: -0.30000001192092896 |
| Sentence sentiment magnitude: 0.30000001192092896 |
| Sentence text: The footprint for a solar or wind farm to produce the same electricity as a dual-unit nuclear power plant is exponentially higher. |
| Sentence sentiment score: -0.20000000298023224 |
| Sentence sentiment magnitude: 0.20000000298023224 |
| Sentence text: Also, the waste associated with solar panel technology is hazardous. |
| Sentence sentiment score: -0.800000011920929 |
| Sentence sentiment magnitude: 0.800000011920929 |
| Language of the text: en |
| Document sentiment score: -0.5 |
| Document sentiment magnitude: 3.200000047683716 |
| Sentence text: @Kahil @AOC Silly response... |
| Sentence sentiment score: -0.4000000059604645 |
| Sentence sentiment magnitude: 0.4000000059604645 |
| Sentence text: I'm saying it very likely won't happen to anyone. |
| Sentence sentiment score: -0.6000000238418579 |
| Sentence sentiment magnitude: 0.6000000238418579 |
| Sentence text: Do you really think people have already received a invoice for electricity used 5 days ago? |
| Sentence sentiment score: -0.699999988079071 |
| Sentence sentiment magnitude: 0.699999988079071 |
| Sentence text: Not at all likely.. |
| Sentence sentiment score: -0.699999988079071 |
| Sentence sentiment magnitude: 0.699999988079071 |
| Sentence text: these are probably account views, which mean little to nothing at this point. |
| Sentence sentiment score: -0.6000000238418579 |
| Sentence sentiment magnitude: 0.6000000238418579 |
| Sentence text: Think it through.. |

Sentiment analysis can be compared to the VADER sentiment analysis in that they both evaluate sentiment without identifying the subject of that sentiment. Google's API reports two real numbers: Sentiment Score and Sentiment Magnitude. The Score will be a number between -1.0 and 1.0 with positive numbers indicating a positive sentiment. The Magnitude is a real number greater than 0, with no theoretical upper range.

In our investigation, the two entities with the highest Magnitude of sentiment associated with them were consistently ERCOT and Ted Cruz. However, amongst posts that did not specifically mention those two entities, the scores were more difficult to interpret (Table 3).

*Table 5 – Google Entity Analysis*

| |
|---|
| Representative name for the entity: solar |
| Entity type: OTHER |
| Salience score: 0.1947712004184723 |
| Mention text: solar |
| Mention type: COMMON |
| Representative name for the entity: plants |
| Entity type: LOCATION |
| Salience score: 0.17020247876644135 |
| Mention text: plants |
| Mention type: COMMON |
| Mention text: plants |
| Mention type: COMMON |
| Representative name for the entity: TX |
| Entity type: LOCATION |
| Salience score: 0.156732082366943336 |
| mid: /m/07b_1 |
| wikipedia_url: https://en.wikipedia.org/wiki/Texas |
| Mention text: TX |
| Mention type: PROPER |

When using the Entity analysis (Table 4), the Google API reports this information as a list of entities, with some associated information. If the entity has a web presence that is easily discoverable by Google, like a Wiki page, the API will return that information along with the quantitative data about the post.

And finally, the Entity Sentiment analysis seems to provide value by showing how salient the given entity is when determining sentiment of the Tweet (Table 5).

Further investigation is required to determine if there is an appropriate method using algorithmic methods that might succeed in providing further insight through clustering or classification.

*Table 3 – Google Entity Sentiment Analysis*

| |
|---|
| Representative name for the entity: environment |
| Entity type: OTHER |
| Salience score: 0.06775835156440735 |
| Entity sentiment score: 0.0 |
| Entity sentiment magnitude: 0.0 |
| Mention text: environment |
| Mention type: COMMON |
| Representative name for the entity: @tedcruz |
| Entity type: OTHER |
| Salience score: 0.05958884581923485 |
| Entity sentiment score: 0.0 |
| Entity sentiment magnitude: 0.0 |
| Mention text: @tedcruz |
| Mention type: PROPER |
| Representative name for the entity: electricity |
| Entity type: OTHER |
| Salience score: 0.0452093742787838 |
| Entity sentiment score: 0.0 |
| Entity sentiment magnitude: 0.0 |
| Mention text: electricity |
| Mention type: COMMON |

But at this point rather than depend on an outside source of training data. It made more sense for our qualitative investigation to switch to a more traditional grounded theory workflow of human interpretation and coding of the data. The NLP provided some basic initial insight as to how folks were talking about the events in Texas, but the NLP investigation to this point was not intended for any kind of theory building, it was intended as investigatory, and we believe it fulfilled that purpose.

## CODING

Pursuant with a bottom-up methodology, we then proceeded to code the individual Tweets, Interviews and Media articles. We used ATLAS.ti [39] as the principal tool for coding. We used Python to parse the data from the JSON returned by the Twitter API and exported each Tweet as an individual text document that lends itself to coding in ATLAS.ti.

We follow the academically accepted grounded theory methodology [28][40] of building up from the bottom in terms of identifying specific terms then grouping those into categories, from which we can extract themes and possibly build theory. The initial process is commonly referred to as "Open" coding, and the next level is often referred to as "Axial" coding, and finally one would get to "Theoretical" coding [28].

This is an iterative process and requires the researcher to be guided by the data while interpreting it. After investigating the Tweets with the NLP, and by applying a "stop list" in ATLAS.ti to get a count of the most frequent words, we arrived at a list of Open codes (Table 6). Then we placed those codes into groups, which can be thought of as Axial codes (Table 7).

As noted, we applied some of the Hermeneutic approach at this stage as we tried to move from the specific to the general and back again. In looking at the Axial codes, the researcher is tasked with extrapolating theory from the codes. In our case, we identified two Theoretical codes: Trust and Time – and both should be thought of as commodities.

We plotted the codes and the content of the articles in Atlas.ti, where they are referred to as networks. This is a very manual process and requires the researcher to examine the data while building out the graph. It was this aspect that we found to be similar in approach to a Hermeneutic Circle, whereby the researcher is continuously cycling between a local and a more general perspective. We have included four networks in the appendices.

*Table 6 – Open Codes*

| @ERCOT_ISO | Angry | Anti-Renewable |
|---|---|---|
| Anti-Republican | ARS | Biden |
| Bill Magness | Blame | Brown/Black-Outs |
| California | Clean Air | Coal |
| Cold | Competition | Electricity |
| Emergency | Energy Consumers | Energy Producers |
| Energy Sellers | Environment | ERCOT |
| Explanation | Extremes | Facts |
| Fail | Federal Assistance | Fiber |
| Fixed Price | Food Prices | FOX News |
| Free-Market | Frozen | Generators |
| GOP | Greedy Corporations | GregAbbott_TX |
| Location Report | Mid Terms | Mumbai |
| National Power Grid | Natural Gas | Non-Texan |
| Nonprofit | Not Connected to Nat'l | Nuclear |
| Oil | Oil Pipeline | OTF |
| Political Spin | Politics | Power Companies |
| Power Outage | Prediction | Price Gouging |
| Railroad | Reliable | Relief Suggestion |
| Representative | Roads | Sarcastic |
| Shortfalls | Solar | Southwest Power Pool |
| State Legislature | Ted Cruz | Texas / Texan |
| Tik-Tok | Time Frame | Trump |
| Unreliable | Utility Bill | Weather Report |
| Widespread | Wind | |

*Table 7 – Axial Codes*

| Descriptive |
| --- |
| Emotions and Attitudes |
| Energy or Utils |
| Environmental |
| Location-istic |
| Media |
| OTF - Opinion True False |
| Political |
| Texas v Everywhere Else |
| Unrelated and Spam |

# EMPIRICAL CONCLUSIONS

Over the course of the investigation, our understanding of the situation has developed such that we believe we can make some interesting observations that contribute at both a theoretical level, and from a practical decision-making and actionable perspective for stakeholders thinking of locating in Texas or those in Texas who rely on an informed understanding of electrical production and consumption.

While we don't want to "blame" anyone or anything, we believe that by documenting our findings in an academically rigorous manner, the conclusions drawn can be placed within their ethnographic context and should therefore be generalizable.

Consistently, a primary concern of many interviewees (familiar with the events in Texas or not), Tweets, and media articles is *where the blame should fall*. Almost universally, they all felt that by coming to what they believed was a factual understanding of events and causes, that they could use that to judge the value of their news sources.

We found that this was often a circular process of reading news that provided a certain viewpoint, which the reader then takes as fact, and from then on, anything that aligns with that primary view is given higher weight. This aspect of bias has been well documented in countless studies and academic journals. Also, well-documented, is that the reader tends to follow knowledge chains that align with their given understanding. Both these ideas are exemplified by the following quote from one of our interviews:

> *I think I definitely trust whatever I find more this way, because I start from things that I know. So, if I trust one thing, then when I go somewhere from that I tend to trust it a little more. Even if it's not directly related.*
>
> *Like I might go from MMA to Fashion and trust that fashion advice more because I got there from an MMA site.*
> -Int #2

Where our analysis is perhaps more specific than others is that we found that users quite often dismissed an opposing viewpoint as a "waste of time" – and we take this phrase literally. Users choose not to read opposing viewpoints not because of a political or any other form of bias, but because they don't want to waste any time.

This was a common theme, and merits repeating - users don't read, look at, or use another platform because they think it's a waste of their time resources, not because they don't want to understand an opposing political, economic, racial, sexual, or other perspective.

And this in turn leads, perhaps to an understanding of why the term "Fake News" has such power and traction. Users really want to be able to trust their source of news. This is their primary concern. Not because they want to be biased in their thinking – quite the opposite. Our analysis shows that consumers of news across all platforms and sources want to know that when they expend their time resources for the acquisition of "news" that they can trust that news as to how it affects their daily lives, then make decisions about their lives based off that information.

"Fake News" is an inconvenient waste of precious time resources at best and detrimental to their health or well-being at its worst. Anyone who is wasting the users' time is by definition interfering in their daily lives. Whether this is a single Black mother in Detroit, a millionaire white male cattle farmer in Texas, a gym teacher in Atlanta, an immigrant or 3[rd] generation U.S. citizen, gay, straight, or anyone else – and however they value their time, anyone wasting it is hurting them – and by extension, their families and friends.

And the "news" used to be considered so trustworthy and fundamental to an American's understanding of the world beyond their block, street, or town that when that trust is shattered, it feels like a fundamental assault on America. That waste of time is not just an inconvenience, if the "news" is not telling the "truth" they must have a nefarious reason.

It is beyond the scope of this paper to explore the plethora of reasons that users then attribute as causes for an entity (person or organization) not telling the "truth." We have, however, examined some primary thematic and philosophical reasons presented in our window of events in Texas. Ethnography allows us to generalize from the observations within that limited window. So, while we can't explore all the possible motivations of every user of every platform, in every case – we can attempt to do so within a limited corpus of one week, 482 Tweets, 18 interviews, and 20 media articles.

And we can use this lens of *trust in the source* as a fundamental unit of narrative. Similar to the approach employed by Webb and Mallon [41], we employed coding very early in the process, and have allowed our understanding to be guided by those codes.

The analysis shows that many people do not get their news from "news sources" at all, but the sentiments and even quite often the specific phrasing is often repeated.

Our analysis shows that there is a connection between the socio-political context of the codes, and the quantitative information embodied by U.S. Census data [42], Neilsen [43] ratings, and election data.

We have quantitative statistical data that confirms some aspects of our qualitative analysis - we can say that if a person lives in a given county in west Texas, in a qualitative sense, it is very likely that they get their news from FOX or someone who does. Whereas if they live in Austin, it is likely they don't get their news from FOX or someone who does.

Further, one can qualitatively predict the source of a user's news, based on the political or technical understanding of the events in Texas. If someone thinks that renewables are to blame, its likely they got their news from FOX news. If someone thinks combustibles (oil, gas, coal, nuclear) are to blame, it's likely they did not get their primary news from FOX.

We find that there is, however, a fundamental difference in types of viewers across both demographics, but apparently far more so amongst FOX, OANN, and Newsmax for opinion-based information. Whereas CNN and MSNBC's ratings seem to go down when they insert too much opinion, FOX's response to a rating dip was to replace news with more opinion.

More detail is provided below, and illustrated with the use of knowledge graphs, but as mentioned before, these viewers tend to seek reinforcement of an ideology, rather than new information. They tend to think they understand as much as they need to know, and are looking for reinforcement to confirm their opinion, thereby making it fact. They believe that if enough people believe the same thing, that is evidence of "truth" – very much an interpretivist realism.

This is to be contrasted by viewers of all media, including FOX, OANN, and Newsmax who are looking for "factual" information. Within this fact seeking group, we found evidence of two types of philosophies – 1) Those who are looking for facts that support their opinion 2) Those looking for facts regardless of content.

The first group is looking for facts that support their opinion, and if they don't get them, they discount the value of the news source as a waste of time. The second group is looking for facts that are important to their current daily situation. They may get a lot of information, but if the information is not relevant to their life, that too will be considered a waste of time. Both groups are therefore positivistically evaluating the news.

The first group is far more likely to take the information personally, because it is tied to their intellectual identity of themselves. Whereas the second group is more likely to express their wasted time as frustration with not being able to find the information they want.

We think there is a clear division to be made between two types of news outlets with regard to the amount of opinion and/or user generated content versus investigative journalism. Both CNN and FOXNews (as well as many others, but CNN and FOX are exemplary of their respective situations) rely on opinion, and/or user generated content. CNN focuses more on the voices of their viewers - they often have streams of Tweets live, or many quoted Tweets in an article - as evidence of whatever point they are trying to make.

FOX and MSNBC simply insert opinion from their hosts, writers, or advertisers. All of those can be contrasted with the longer articles from ARS Technica (as well as many others).

We are able to segment an article by using a third Axial Code Group (Table 8) to try to identify opinion, "fact," and the verity of each statement when compared with our empirical understanding. In the article depicted in Figure 6, we see that only four of the seven segments meet the standards to be described by users as "Facts." Two segments are opinion, that fallaciously link current events to other political or ideological agendas. And finally, the article ends with an opinion that appears neutral, but in reality, re-frames the argument in an illogical and philosophically unsound way.

Contrast this with the ARS Technica article that contains no opinion but does include a forecast: based on ERCOT's current estimates, demand would outstrip supply later that afternoon - and this was in fact what happened. They were the first outlet that these researchers were able to find that accurately described the empirical situation while offering advance warning for consumers.

*Table 8 – OTF Code Group*

| |
|---|
| OTF - T - "Fact" based statement that matches coder's empirical understanding |
| OTF - F - "Fact" based statement that does not match coder's empirical understanding |
| OTF - OT - Opinion matches coder's empirical understanding |
| OTF - OF - Opinion does not match coder's empirical understanding |
| OTF - TW - Weather |
| OTF - O - Opinion no "Facts" |
| OTF - OTBL - Opinion matches coder's empirical understanding but is logically deceptive |

*Figure 6 – Mainstream News article diagram*

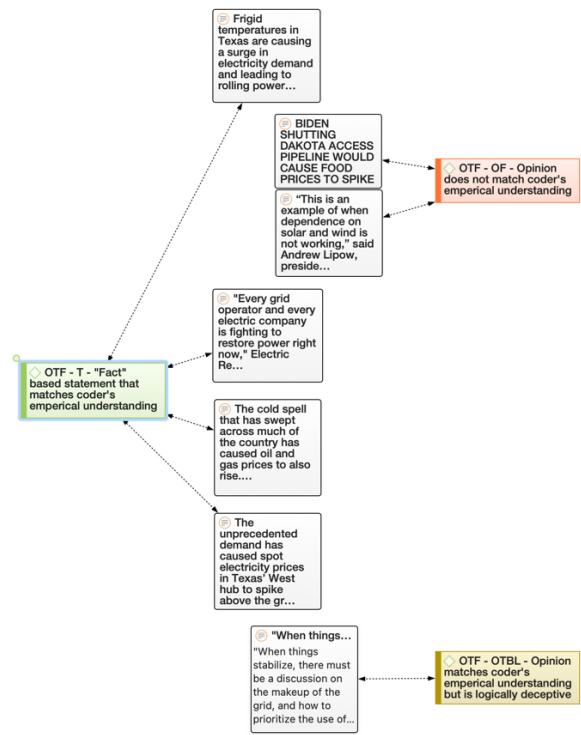

In a qualitatively descriptive sense, what distinguishes these articles is the amount of primary content collected by the reporter, and a delivery that may have political implications, but is not written because of those implications. Essentially, this distinction mirrors the way Twitter users, and interviewees described their ideas of "Traditional Mainstream Media."

And according to the network graphs, the Twittersphere was equally capable of delivering relevant information to users as the FOX or CNN article, but it was not as efficient as the ARS Technica article.

Yet the readership and viewership of the outlets appears to lean the opposite direction. The article with the highest proportion of opinion has the highest readership. The article with the lowest percentage of opinion has the smallest distribution. This indicates that consumers of news, from all ranges of the political spectrum, are not recognizing their own bias. Consumers say they want "facts," but they are far more interested when those facts are bookended by opinion that matches their own, additional facts are secondary. And facts that contradict the opinion are discarded entirely.

It appears that based on the Tweets, and Media sentiment, and interviews with those who identified as familiar with the events in Texas, ERCOT has been the focus of significant blame. This is evidenced in our network graph and makes sense in many ways. However, academic discipline and philosophical rigor requires that we challenge this assumption based on our understanding of the empirical data.

From the consumer's perspective, no matter whether they get their electricity from combustibles or from renewables, they get it via ERCOT. In practice, there is usually a local energy provider between the consumer and ERCOT, and interestingly, while few consumers seem to have heard of ERCOT before the outage, many did not associate the outage with their local utility, but with ERCOT.

*I never heard of ERCOT until the power went out, but man, they sure screwed up!*

Our research shows that there is an organic and natural function of market regulation provided by ERCOT, and that within the boundaries of their charter, they were operating efficiently and transparently as a market for energy.

In the same way that markets in Alaska, New England and the Gulf states provide fishing and the associated ports, markets, facilities and other infrastructure a means to regulate the price of the fish delivered, so too can the local ERCOT market serve as an effective means of competitive pricing. This competitive pricing can deliver value to the consumer as well as the producer.

But like the fishing markets, or any other free-market, issues of cost may be hidden. Concerns by economists to distinguish between the measurable costs and the hidden costs of utility-scale electricity were addressed in the 2010 book commissioned by Congress and investigated by the National Science Council of the National Academy of Sciences [44]:

*The U.S. Congress directed the U.S. Department of the Treasury to arrange for a review by the National Academy of Sciences to define and evaluate the health, environmental, security, and infrastructural external costs and benefits associated with the production and consumption of energy — costs and benefits that are not or may not be fully incorporated into the market price of energy, into the federal tax or fee, or into other applicable revenue measures related to production and consumption of energy.*

Per the economic laws of capitalism, decisions of safety and/or reliability are driven by the consumer's willingness to accept a given price to accommodate a given result. A free market allows consumers to determine whether to pay more or less for a given level of service.

The fundamental breakdown in Texas was between the predicted need, and what the real demand was. While the free-market system worked as an effective means to determine a price for any energy that was available, it failed to provide any assurance that sufficient providers would choose to winterize their equipment for the most extreme of events. So, the market performed as expected, the price of electricity skyrocketed as demand soared relative to the available supply.

In fact, this sudden increase in price would later become the subject of a second common concern. For many electric users whose power remained on, their experience was one of seeing their price for electricity soar to thousands of times the normal rate in a matter of hours meaning their electric bill would be hundreds of times the usual bill. 1 hour x 1000 = 41.67 days of usage being billed for every hour they used their power to heat their home or power a factory or hospital. In fact, for many this multiplier was well above a 6000x multiplier that had been put in place as a "safety" limit.

Under normal conditions, and as has happened every winter since 1995 when ERCOT

was formed 26 years ago, the rising price of power would have induced more providers to bring their facilities online, or ramp up production, but the cold made this an impossibility. According to NOAA, this was the coldest it had been in 30 years.

And as demand increased, unsustainable electrical loads were placed on any junction boxes and power lines that had not been physically damaged in the storm. This in turn caused a snowballing effect - as it got colder over the next two days, increased demand created increased load on the lines – again further increasing the need to preemptively shut down power generators to reduce the strain and avoid overload or fire.

So, in response, instead of being able to use rolling 4-hour "brown outs" to spread the interruptions and allow users to recharge batteries and other devices in between brown outs, the utility companies were suddenly forced to shut off power completely to large swaths of the population – a "black out".

For the utility providers, it was not just a question of getting the power lines back up, then turning the switches back on. There is an order that needs to be followed when powering up the grid to avoid simply tripping the breakers again. Unbeknownst to them at the time, there were unforeseen issues emerging from the production side.

The pipes carrying the oil and natural gas had gotten so cold, the contents therein were freezing. Thermal / Hydro power was supposed to provide 83% of the estimated power to satisfy demand during the winter (Figure 2), yet collectively was operating at 50-60% efficiency, satisfying only around 40-50% of the real demand (Figure 3).

Coal plants that had been expected to continue operation found their operations intermittently halted by unforeseen failures in tertiary equipment caused by the extreme cold.

Some wind turbines froze in place, and some solar was covered in snow. Yet, wind and solar were actually producing more than had been predicted – approximately 16% of the predicted demand – whereas it had been assumed they would only need to provide approximately 8% during winter months.

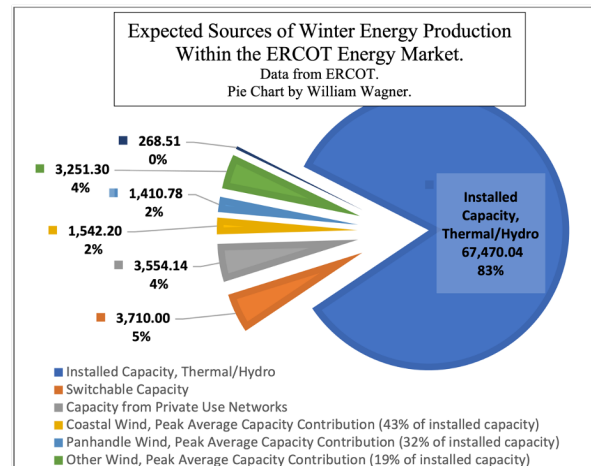

Figure 7 - Expected Sources of Winter Energy

One of the four nuclear plants was taken off the grid to avoid electrical overload on its delivery lines, reducing the amount of nuclear generated energy by 25%. But since nuclear power accounts for a small percentage of overall power generation in Texas, the overall effect of the lost power was not significant. And there was never any danger of any nuclear disaster.

However, as mentioned above, the power grids need to be brought online in a given order to prevent overloading the system while starting up. The result of the reductions in power from the production side meant that certain locations that needed to come online before others were unable to come online when required to do so. So, although the nuclear plant was ready to come back online, it was forced to wait while other utilities restored their operations so that loads would be balanced across the wires.

So, all of a sudden, residents that had been told only hours earlier that they should expect rolling brownouts, were suddenly informed that the brownouts would be blackouts, and that there was no estimated time for repair.

At the same time, temperatures remained far below normal in many places in Texas. The conflux of technical issues created a situation where the only real possibility for a return to normalcy was dependent on the weather warming back up. Some emergency power was returned as possible, but a genuine return to reliable power would be four days away for approximately 10 million people.  And during those four days, approximately 100 people in Texas died from hypothermia and other cold-related deaths like carbon-monoxide poisoning from running a car or truck in the garage for power inside the house.

Many consumers associated the outage with their elected officials in terms of regulatory enforcement.  There are five basic axial codes within which we can place the blame:  Local Utility, ERCOT, State Government, Federal Government, No Blame (unpredictable freak weather event).

While Ted Cruz received much bad press for going to Cancun during the outage, it focused on the fact that he was escaping a bad situation instead of helping and taking a constructive leadership role to fix it.  It should be made clear that Ted Cruz is one of two U.S. Senators from Texas who serve in Washington D.C. and is not part of the Texas State Legislature or the committee(s) that oversee ERCOT and/or regulate the Texas energy market.

There was little discussion as to whether his policies, or policies for which he had voiced support, had contributed to the outage, only that his failure to pitch in to help those in need during the crisis was not a good image.  There was a great deal of concern about the fact he had left his dog behind in a frozen house.

When he took his trip to Cancun, his name rises dramatically to the top of the list in terms of entity recognition and magnitude of sentiment. Texans clearly felt this did not exemplify what it meant to be a "good" Texan.  They felt betrayed by this man who very closely associates himself in public address and political concern as being a "real Texan".  Again, not because they disagreed with his free-market approach to the energy market in Texas, but they were angered by his privilege that allowed him to jump on a plane with his family and a few friends and fly away to Cancun, while other Texans were dying in the cold.

## FUTURE RESEARCH

We believe there is opportunity for further research in the use of an emerging technique known as Knowledge Graphing.  This technique seeks to link ideas and entities via semantic means rather than a hierarchical structure. In many ways, it is similar to the network diagrams created by ATLAS.ti but distinguishes itself in that Knowledge Graphs may take the form of AI generated content stored in a graphical database.  This may provide opportunity for both a "Semantic World Wide Web" as well as methodological considerations for Qualitative research data collection and analysis.

Secondary factors like price, availability, device compatibility, friends using a platform, and other factors were not explored quantitatively, but in our qualitative analysis, they were all secondary to the two primary concepts of Trust and Time.  In other words, consumers of the news seemed more inclined to want to pay for accurate news, than get unreliable news for free.  And they would travel or change platforms if they thought they would get more value from their news.  And even friendships and familial relationships have shown significant strain in the last few years because of differences in news source.

## CONCLUSION

Humans have an innate thirst for knowledge.  From the earliest days, homo sapiens have understood that their survival depends on learning what things can hurt or kill them, versus those that can help them.

Information exchange is fundamental to society, and society is in turn responsible for the way that information is created and shared.  It is a circular arrangement that both defines and is defined by its participants.  Our understanding of that information flow and the knowledge and power contained therein leads us to believe that very little current "News" can be thought of as "Trusted."

Although consumers profess to desire "unbiased" "facts", they are unwilling to tune in to any source that contradicts their understanding of the world in their vision.  It is clear from our analysis that any bias brought by a viewer is not viewed as a negative by that consumer.  The viewers see their bias as informed opinion that they are more than happy to use as a litmus test.

Any attempt to overcome those biases cannot rest on ideological grounds. It must be a clean start and must never contain any opinion. However, a purely emergency channel that provides only verified information may not be responsive enough to provide the type of information many users are looking for in an emergency.

Along with the obvious official announcements that users would need, we found that consumers who turn to news sources in an emergency consider other factors to maintain interest – they want regular updates, and they want more than just numbers, locations, and times.

This dichotomy of "boring" factual information with the more exciting styles of human interest, Twitter feeds, or opinion is then affected by the fundamental forces of economics. In order for a station to remain on the air, or to return profit to its shareholders, it attempts to maximize its viewership. It caters to its audience. In economic terms, one could say they understand their consumer and are delivering the product their consumer values. Philosophically, this is not fundamentally a "bad" thing.

Logically, the greater the reach of an emergency information channel, the more effectively it can spread helpful information during a disaster. Unfortunately, the converse is also true. And when political or financial interests pollute the discourse regarding an emergency, it creates cataclysmic effects that damage not only the individuals provided with false information, but also the very foundations humanity needs to survive.

All civilizations and governments depend on the participation of their citizens to act in support of that civilization. Whether that support is forced or given is what defines modern political discourse. What our research shows is that intentionally spreading information that is unverified or attempts to favor one political view during an emergency has widespread secondary effects that often lead to violence.

The power of a statement like "You see you have no electricity - it's because group X did action Y," is that it ties real-world experience to an ideology in a way that the viewer has already chosen to accept, and now they have "evidence" they can see – a black out – and confirmation of its cause from a "trusted" source. In logical terms, we recognize that as an unfounded warrant, but this level of dissection is rarely applied by consumers in our research.

It is important to note that an unfounded warrant can still be correct, it just lacks any logical foundation. However, if the warrant is not accepted, then any arguments based on that warrant cannot be proved. So, a warrant requires the consumer and provider to agree on an ideological foundation *before* any information can be exchanged. This is clearly not appropriate if the information being shared is of a current and emergency nature – therefore, underlying political or economic ideology should have no place on an emergency news channel.

Further, not only should there be no opinion, but there should also never have been any opinion. Any source that intends to be considered "Trustworthy" "news" should start out with that intent, and never mix opinion and news. While there are certainly profound reasons why opinion and political commentary should not be excluded from the national current events reporting, if an organization wants to increase the perceived trust of its viewers, *any opinion will reduce that trust*.

The concept that any opinion, whether the viewer agrees with the opinion or not - reduces trust in the source is fundamental. This seems to indicate that many viewers understand their own biases and are aware of the spins the media or Twittersphere applies, then they use a comparison of their own bias with that of the outlet as a filter to separate "fact" from opinion.

There is a particular jargon that each of the consumers is comfortable with, and this reduces the effort required to interpret the news. So, although they prefer to have their "news" match a particular ideology, they also recognize that ideology to be different from "facts."

It is this tension that we believe has caused a downward spiral in Americans' faith in the mainstream media. This trait that even those who agree with the opinion will still have less faith in that news outlet because they included any opinion at all is inherent in any information system that attempts to mix opinion with empirical observation without explicitly detailing any underlying philosophy and methodology.

This need for information is so innate, that despite any evidence to suggest otherwise, a significant number of Americans report a lack of faith in the electoral system because their news source has suggested there was a problem with the 2020 U.S. Presidential elections. And this in turn may have been a contributing factor to the violence on January 6, 2021, at the U.S. Capitol – which many consider a failed insurrection.

This research has demonstrated two fundamental aspects of humanity's need for information: 1.) Trust in the source and 2.) Time as a resource. When either is diminished, many users will seek an entity to blame, quite often based on an ideology rather than philosophically sound logic.

It is not without some sense of irony that we point out this inherent flaw in the information systems we think of as news. Academic institutions and peer-reviewed publications face similar challenges in both the distillation as well as distribution of scientifically sound information. And we hope this paper demonstrates both internal and external validity, sufficient logical and philosophical rigor, and the requisite thick description that this ethnographic account may be generalized well beyond the scope of the events of one week in Texas.

# BIBLIOGRAPHY

[1] "Assessing the U.S. Climate in February 2021," *National Centers for Environmental Information (NCEI)*, Mar. 05, 2021. http://www.ncei.noaa.gov/news/national-climate-202102 (accessed Apr. 19, 2021).

[2] Melissa Macaya, "Snow and ice storms: Millions without power as winter weather blasts the US," *CNN*. https://www.cnn.com/us/live-news/snow-ice-storms-updates-02-15-21/index.html (accessed Apr. 25, 2021).

[3] P. Godfrey-Smith, *Theory and reality: an introduction to the philosophy of science*. Chicago: University of Chicago Press, 2003.

[4] C. Geertz, *The interpretation of cultures: selected essays*. Basic Books.

[5] P. Ricœur, *History and truth*. Evanston [Ill.] : Northwestern University Press, 1965. Accessed: Apr. 10, 2021. [Online]. Available: http://archive.org/details/historytruthessa0000unse

[6] M. D. Myers and L. W. Young, "Hidden agendas, power and managerial assumptions in information systems development An ethnographic study," *Information Technology & People*, vol. 10, no. 3, pp. 224–240, 1997, doi: http://dx.doi.org.ccl.idm.oclc.org/10.1108/09593849710178225.

[7] J. Davis and A. Stern, *AN ETHNOGRAPHIC STUDY OF SLA ENACTMENT*. 2015. doi: 10.1057/ejis.2010.51.

[8] M. Zachariadis, S. Scott, and M. Barrett, "Methodological Implications of Critical Realism for Mixed-Methods Research," *MISQ*, vol. 37, no. 3, pp. 855–879, Mar. 2013, doi: 10.25300/MISQ/2013/37.3.09.

[9] J. Kotlarsky, H. Scarbrough, and I. Oshri, "Coordinating Expertise Across Knowledge Boundaries iin Offshore-Outsourcing Projects: The Role of Codification," *MIS Quarterly*, vol. 38, no. 2, pp. 607-A5, 2014.

[10] Texas, D. E. Simmons, G. P. Finlay, C. W. Raines, H. P. N. Gammel, and Coahuila and Texas (Mexico : State), *The laws of Texas 1822-1897: Austin's colonization law and contract; Mexican Constitution of 1824; federal Colonization law; Colonization laws of Coahuila and Texas; Colonization law of state of Tamaulipas; Fredonian declaration of independence; laws and decrees, with constitution of Coahuila and Texas; San Felipe convention; journals of the consultation; proceedings of the General council; Goliad declaration of independence; journals of the convention at Washington; ordinances and decrees of the consultation; declaration of independence; constitution of the republic; laws, general and special, of the republic; annexation resolution of the United States; ratification of the same by Texas; Constitution of the United States; constitutions of the state of Texas, with all the laws, general and special, passed thereunder, including ordinances, decrees, and resolutions, with the constitution of the Confederate states and the Reconstruction acts of Congress.* Austin: The Gammel book company., 1898. [Online]. Available: //catalog.hathitrust.org/Record/010476684

[11] J. C. Townes and Z. T. Fulmore, "Review of The Laws of Texas, 1822-1897," *The Quarterly of the Texas State Historical Association*, vol. 5, no. 3, pp. 258–262, 1902.

[12] E. R. Schlereth, University of Texas at Arlington, G. D. Saxon, and S. W. Haynes, *Contested Empire : Rethinking the Texas Revolution*, vol. First edition, no. number forty-six. College Station: Texas A&M University Press, 2015. [Online]. Available: http://search.ebscohost.com/login.aspx?direct=true&AuthType=sso&db=nlebk&AN=1023159&site=ehost-live&scope=site&custid=s8438901

[13] "TSHA | Texas Revolution." https://www.tshaonline.org/handbook/entries/texas-revolution (accessed Apr. 24, 2021).

[14] "Texas Revolution | Causes, Battles, Facts, & Definition," *Encyclopedia Britannica*. https://www.britannica.com/topic/Texas-Revolution (accessed Apr. 24, 2021).

[15] "TSHA | Texas Declaration of Independence." https://www.tshaonline.org/handbook/entries/texas-declaration-of-independence (accessed Apr. 24, 2021).

[16] F. Merk, *Manifest destiny and mission in American history : a reinterpretation*. New York : Vintage Books, 1966. Accessed: Apr. 24, 2021. [Online]. Available: http://archive.org/details/manifestdestiny00fred

[17] "Presidential Election Results." https://www.sos.state.tx.us/elections/historical/presidential.shtml (accessed Apr. 24, 2021).

[18] C. C. Rister, *Oil!: Titan of the Southwest*. Norman: University of Oklahoma Press, 1949. [Online]. Available: //catalog.hathitrust.org/Record/001115481

[19] "A street scene during oil boom days, Desdemona, Texas," *UTA Libraries Digital Gallery/digitalgallery-beta*, Jan. 17, 2018. https://library.uta.edu/digitalgallery/img/10001994 (accessed Apr. 23, 2021).

[20] "About ERCOT." http://www.ercot.com/about (accessed Mar. 28, 2021).

[21] ERCOT, "Capacity Demand and Reserves Report Dec2020.xlsx." ERCOT.

[22] "Official 2020 Presidential General Election Results," Federal Election Commission, 2020.

[23] "Definition of BLOC." https://www.merriam-webster.com/dictionary/bloc (accessed Apr. 17, 2021).

[24] "U.S. most-watched news network 2020," *Statista*. https://www.statista.com/statistics/373814/cable-news-network-viewership-usa/ (accessed Apr. 16, 2021).

[25] "Nielsen | Audience is Everything." https://www.nielsen.com/us/en (accessed Apr. 19, 2021).

[26] Vanessa Otero, "Download The Media Bias Chart," *Ad Fontes Media*. https://www.adfontesmedia.com/download-the-media-bias-chart/ (accessed Apr. 06, 2021).

[27] Vanessa Otero, "Multi-Analyst-Ratings-Project-White-Paper-Aug-2019.pdf."

[28] M. D. Myers, *Qualitative research in business and management*, 3rd edition. Thousand Oaks, CA: SAGE Publications, 2019.

[29] T. S. Kuhn, *The structure of scientific revolutions,* 2d ed., enl. [Chicago, 1970. [Online]. Available: http://hdl.handle.net/2027/mdp.39015016056130

[30] V. Crapanzano, *Tuhami : portrait of a Moroccan*. Chicago : University of Chicago Press, 1986. Accessed: Feb. 16, 2021. [Online]. Available: http://archive.org/details/tuhamiportraitof00crap

[31] Erik Krogh, "A Grounded Theory Approach Toward Understanding the Motivations for the 'Unsanctioned Use of IT'.pdf."

[32] J. P. Gerlach and R. T. Cenfetelli, "Constant Checking Is Not Addiction: A Grounded Theory of IT-Mediated State-Tracking," *MISQ*, vol. 44, no. 4, pp. 1705–1732, Dec. 2020, doi: 10.25300/MISQ/2020/15685.

[33] "Twitter API Documentation." https://developer.twitter.com/en/docs/twitter-api (accessed Mar. 31, 2021).

[34] G. G. Chowdhury, "Natural language processing," *Ann. Rev. Info. Sci. Tech.*, vol. 37, no. 1, pp. 51–89, Jan. 2005, doi: 10.1002/aris.1440370103.

[35] "Natural Language Toolkit — NLTK 3.6 documentation." https://www.nltk.org/index.html (accessed Apr. 07, 2021).

[36] C. J. Hutto and E. Gilbert, "VADER: A Parsimonious Rule-based Model for Sentiment Analysis of Social Media Text," p. 10.

[37] S. Elbagir and J. Yang, "Twitter Sentiment Analysis Using Natural Language Toolkit and VADER Sentiment," *Hong Kong*, p. 5, 2019.

[38] "Cloud Natural Language documentation | Cloud Natural Language API," *Google Cloud*. https://cloud.google.com/natural-language/docs (accessed Apr. 07, 2021).

[39] "ATLAS.ti: The Qualitative Data Analysis & Research Software," *ATLAS.ti*. https://atlasti.com/ (accessed Apr. 07, 2021).

[40] C. Erlingsson and P. Brysiewicz, "A hands-on guide to doing content analysis," *African Journal of Emergency Medicine*, vol. 7, no. 3, pp. 93–99, Sep. 2017, doi: 10.1016/j.afjem.2017.08.001.

[41] Queen's University of Belfast, B. Webb, B. Mallon, and Queen's University of Belfast, "A Method to Bridge the Gap between Breadth and Depth in IS Narrative Analysis," *JAIS*, vol. 8, no. 7, pp. 368–371, Jul. 2007, doi: 10.17705/1jais.00134.

[42] "https://data.census.gov/cedsci/." https://data.census.gov/cedsci/ (accessed Apr. 18, 2021).

[43] "Top 10s: TV Ratings, Video Games, SVOD." https://www.nielsen.com/us/en/top-ten (accessed Apr. 18, 2021).

[44] National Research Council (U.S.), National Research Council (U.S.), National Research Council (U.S.), National Research Council (U.S.), and National Academies Press (U.S.), Eds., *Hidden costs of energy: unpriced consequences of energy production and use*. Washington, D.C: National Academies Press, 2010.

# APPENDIX A

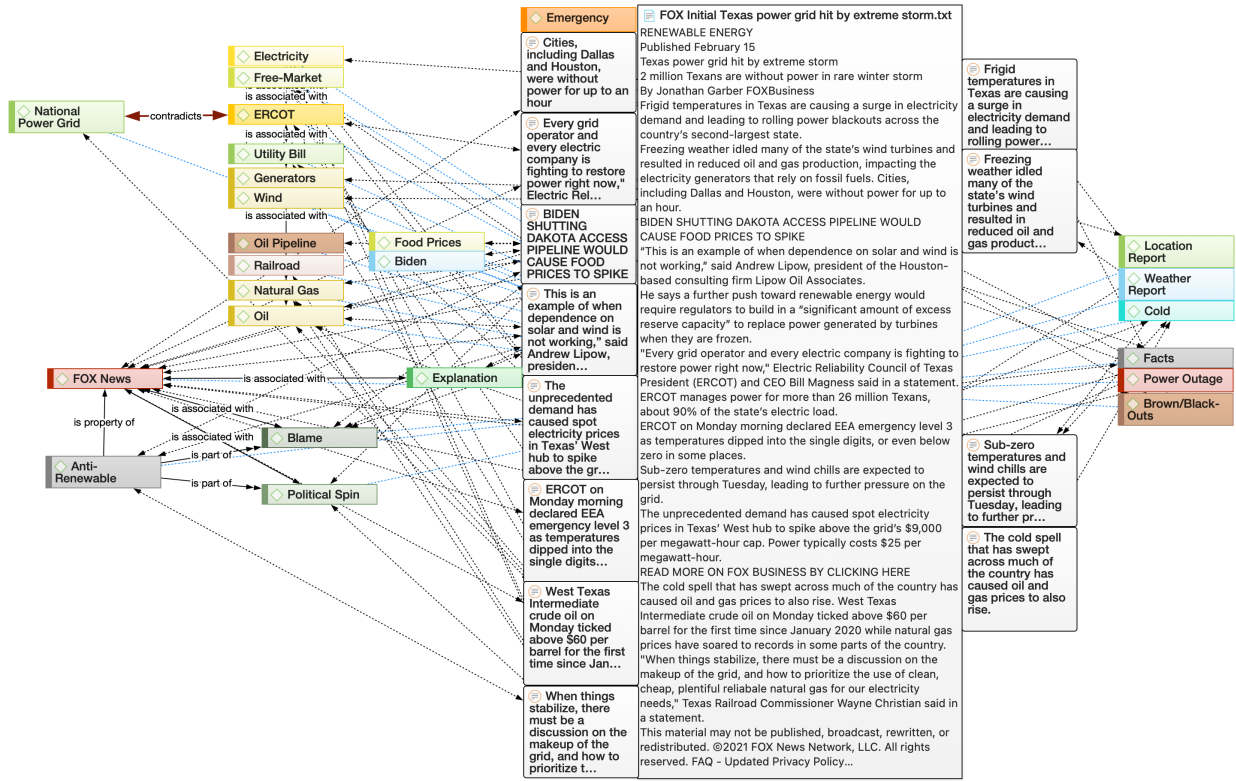

# APPENDIX B

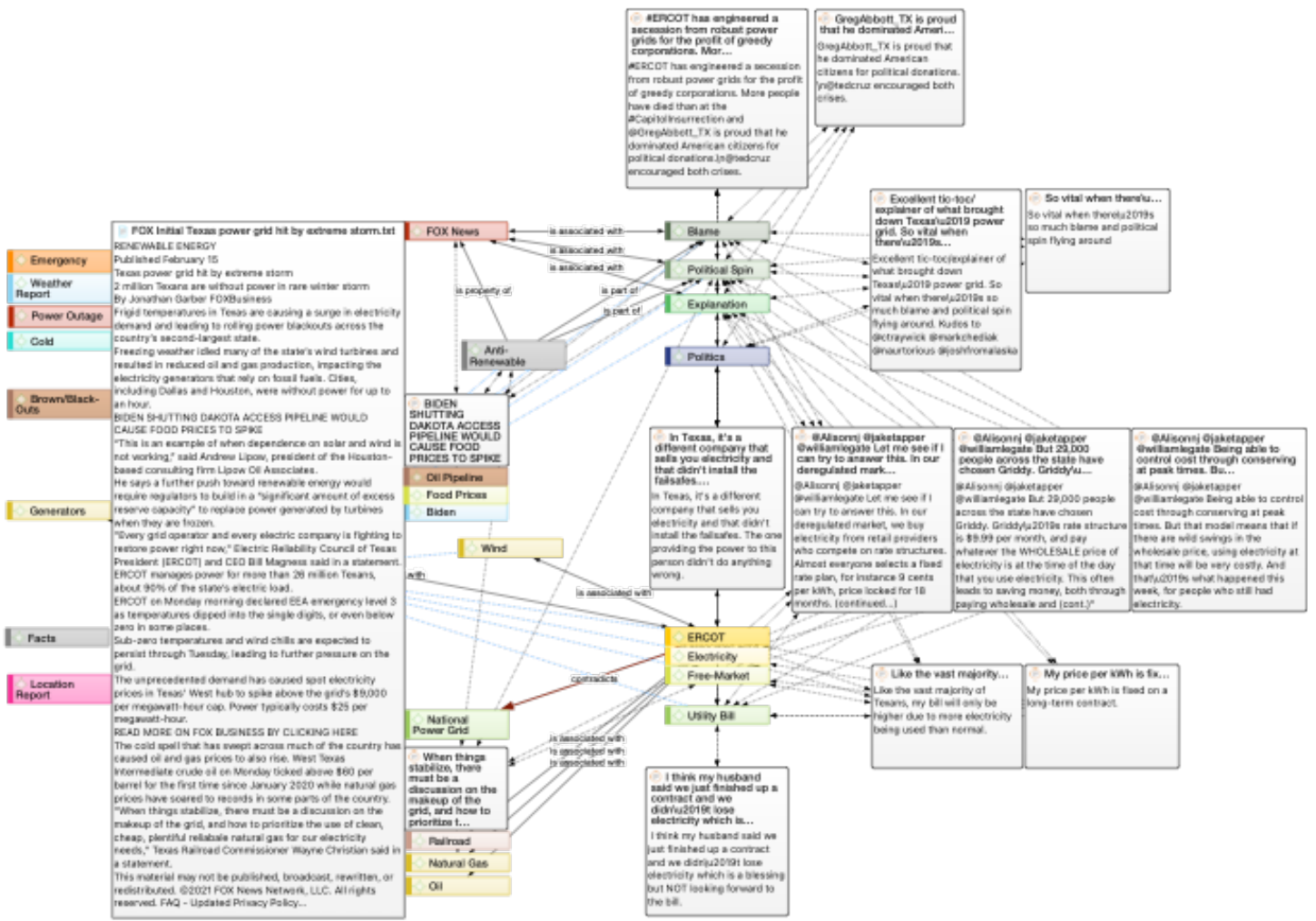

APPENDIX C

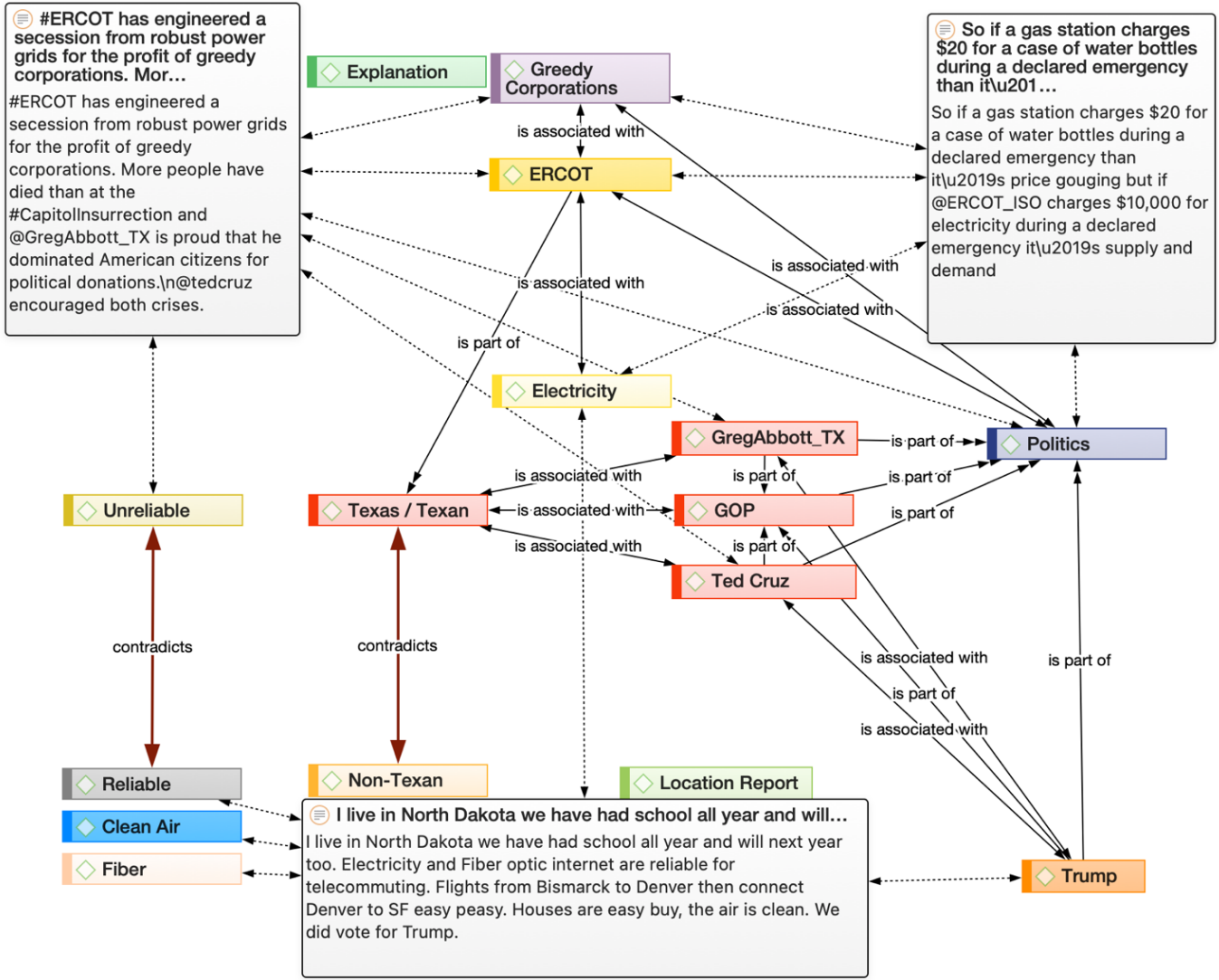

# APPENDIX D

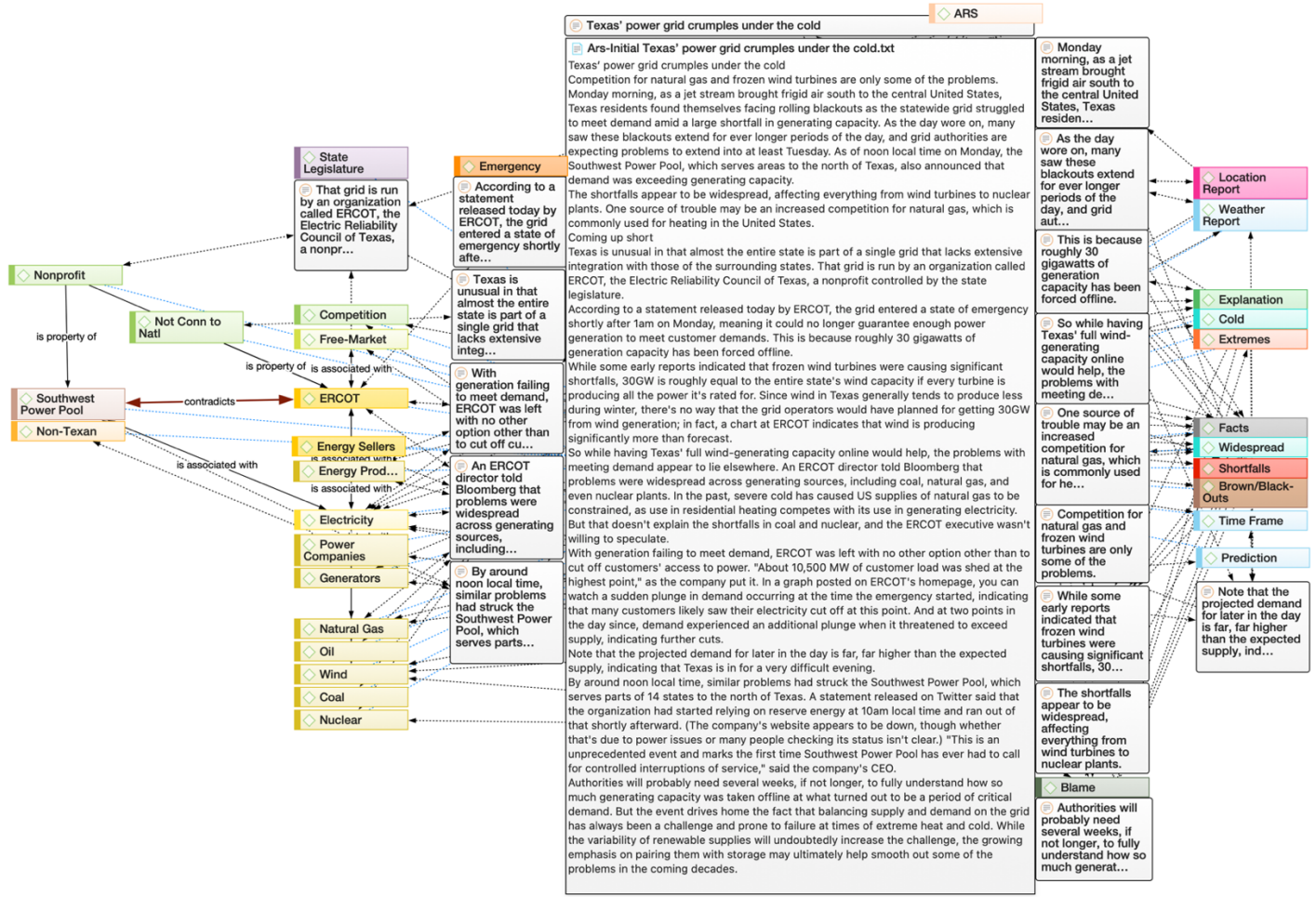